\documentclass{ws-ijmpd}
\usepackage[utf8]{inputenc}
\usepackage{amsmath}
\usepackage{amsfonts}
\usepackage{amssymb}
\usepackage{graphicx}
\usepackage{physics}
\usepackage{caption}
\usepackage{subcaption}

\begin{document}

\title{Probing magnetic correlations in anisotropic turbulence }

\author{Bijita Bose}
\address{School of Physics, University of Hyderabad, Gachibowli, Hyderabad, India 500046}
\author{Somdeep Dey}
\address{School of Physics, University of Hyderabad, Gachibowli, Hyderabad, India 500046}
\author{Soma Sanyal}
\address{School of Physics, University of Hyderabad, Gachibowli, Hyderabad, India 500046}
\date{}

\maketitle

\begin{abstract}

{ The topological classes of coherent structures in magnetohydrodynamic turbulence are studied using the irreducible representations of the SO(3) and SO(2) rotation groups. We probe the coherent structures around axisymmetric magnetic fields in the universe. Such axisymmetric fields can be found around filamentary structures as well as cosmic strings. We have
considered a Gaussian and a non Gaussian magnetic field. We find that the field distribution 
leads to different coherent structures for the two cases. Our results show that the 
scattering of particles from the Gaussian structure depends predominantly on two lengthscales.
In the case of the non Gaussian distribution, we can define a third intermediate lengthscale,
similar to the Taylor lengthscale. In both the cases, we find that there is a dominance of the 
polarization anisotropy which indicates that the field gradients are forced into rigid spatial configurations.  In the early universe, this will change the mean free path of the baryons and leptons 
in the coherent structures. Trapping of charged particle as well as their acceleration will be more frequent in the case of a magnetic field with a heavy tailed distribution. The lengthscale of density 
inhomogeneities close to these structures will thus be determined by the magnetic field distribution of the fields.} 

\end{abstract}


\maketitle

\section{Introduction}

Magnetohydrodynamic turbulence plays an important role in the evolution 
of magnetic fields in the universe \cite{grasso,subramanian}. The origin and evolution of magnetic fields in the early universe is linked to the structure formation in the early universe. In recent times, the anisotropy in the turbulence spectrum has been analysed in the literature by applying the SO(3) decomposition. In
this framework, the complete correlation tensor is decomposed into the  different irreducible representations of the rotation group indexed by an integral spin \cite{arad1,rubinstein,biferale}.  The different integral values of the multiscale correlation 
tensor gives us the different modes of the higher order fluctuations.
The strength of the SO(3) decomposition is that the anisotropic turbulent fluctuations can be disentangled by writing out the complete correlation tensor related to the fluctuations 
in terms of it's modes \cite{staicu}. It has been shown that every mode of the general solution obtained from the complete tensor will satisfy the magnetohydrodynamic equations by themselves \cite{arad}. These higher mode  fluctuations are especially interesting in the case of the magnetic field correlations as the amplitude fluctuations give rise to a set of coherent structures such as current sheets, shocks and filaments. 
Coherent structures refer to self organized patterns which emerge from the turbulent motion of the plasma \cite{lemoine,davis}. 
These may be elongated filamentous magnetic fields localized over a certain area (flux tubes/flux ropes) or thin current sheets in the plasma. 
Since these coherent structures are high amplitude and stable structures, they lead to various observable processes in the magnetized plasma \cite{vlahos}. They will affect the scattering of matter particles moving in the plasma. Other than particle scattering, thin current sheets also lead to the 
phenomenon of magnetic reconnection, whereby the magnetic lines of force break and 
reconnect in a different direction. This leads to a release of a large amount of magnetic energy into the plasma. The sudden increase of energy accelerates the charged particles in the plasma leading to several observable effects such as the emission of synchrotron radiation.  Therefore these structures are important in understanding  magnetohydrodynamic turbulence. They have therefore been studied in detail for the turbulent solar wind \cite{vinogradov}.

Though studied extensively for the current universe, there is little understanding of the role of these coherent structures in the evolution of the magnetic field in the early universe. 
In this particular study, we are interested in the axisymmetric magnetic fields 
found in the early universe. In general, the presence of a magnetic field would break the symmetry of the magnetohydrodynamic flow, as the magnetic field itself has a specific orientation. Such flows have been studied in the context of the Bianchi I cosmology \cite{tsagas} in the early universe. Here we are considering the expanding universe with the FLRW metric. We intend to study the magnetic coherent structures for a turbulent flow in an axisymmetric magnetic field. Such fields may be generated close to cylindrical structures such as filaments and cosmic strings. Another source of such fields are the cosmological phase transitions which often gives rise to tangled magnetic fields 
at small lengthscales. 

We probe the magnetic field correlations at various lengthscales using the SO(3) framework. 
The choice of the magnetic field distribution comes from different kinds of ways in which they are generated in the early universe. We 
initially study the  anisotropic  fluctuations with a Gaussian distribution. A Gaussian magnetic field is the most common form of the magnetic field 
expected in the early universe.  However, there are some primordial magnetic fields generated in 
phase transitions which often have a fat tailed distribution. Intermittent magnetic
field generated in a turbulent medium also tend to have a fat tailed distribution. We 
have chosen to model the fat tailed distribution, by a Cauchy distribution. We find 
interesting differences in the geometry of the magnetic structures formed for these 
two different kinds of distributions. This in turn means that the particle scattering 
off these coherent structures will be different. Generally, coherent structures lead 
to accelerated scattering at certain angles and lengthscales while no scattering at other 
angles or lengthscales. This means instead of free streaming particles, we will have 
localized bounce orbits and trapping of ions and electrons. Our detailed probe indicates 
that for Gaussian fields there are only two lengthscales involved in going from a locally 
coherent flow to a more accelerated flow. For the Cauchy distribution, we however have an intermediate lengthscale between the coherent flow and the sharply accelerated flow.

In the early universe, particle diffusion between the electroweak and the quark hadron transition is important as density inhomogeneities are present in the universe during this time. Most of the particles in the 
plasma at this stage are charged particles consisting of the various quarks and the electrons as well as their antiparticles. The density inhomogeneities have a role to play in the subsequent stages of phase transition and nucleosynthesis \cite{kurkiS,sanyal,das}. The presence of these coherent 
structures and their effect on the diffusion of charged particles will 
determine the size and density of some of these inhomogeneities. 
We will mention briefly about these possibilities in the later sections 
of this paper.   

The structure of the paper is as follows. In Section II, we describe the general mathematical SO(3) framework which we have 
used to study the magnetic field correlations.  In section III, we focus on the 
Gaussian magnetic field fluctuations and it's correlations. Section IV is for the magnetic field fluctuations
modeled by the Cauchy distribution. This is a fat tailed distribution and has been 
used previously to describe a field structure where the plasma's transport properties 
deviate from the standard Brownian transport mechanism. Section V discusses the astrophysical implications of our results and finally in section VI, we summarize and 
conclude.

\section{The SO(3) decomposition of the magnetic field correlations}
In rotational turbulence the two point velocity correlation tensor can be used to analyze the anisotropies arising due to the rotational motion \cite{cambon}. Since the rotational motion is 
associated mathematically with the SO(3) group,  the SO(3) decomposition has been used \cite{rubinstein} 
to generalise the representation of the axial symmetry of an anisotropic tensor to an 
arbitrary anisotropy of the turbulent plasma. This method has been applied mostly to the velocity correlation in turbulent hydrodynamics. Let us consider that $V(\vec{r},t)$ is a statistically homogeneous velocity field of a magnetized plasma and $B(\vec{r},t)$ is the 
corresponding magnetic field. We are interested in the magnetic coherent structures so we 
will concentrate on the magnetic field fluctuations and their correlations in this work. 
For the SO(3) formulation, it is convenient to work in the Fourier space and hence we 
will be considering, 
\begin{equation}
{\bf B}({\bf{k}},t) = \frac{1}{(2 \pi)^3} \int {\bf B}({\bf{r}},t) {exp (-i {\bf{k}}. {\bf{r}}) d^3 r}
\end{equation}
Since these are magnetic fields in the early universe, we will consider the magnetic fields to be frozen in with the plasma.  The magnetic correlation tensor is then defined as, 
\begin{equation}
M_{ij} ({\bf{k}}) = \langle {b_i}({\bf{k}}){b_j^*}({\bf{k}}) \rangle
\end{equation}  
Here, ${b_i}(\bf{k})$ is the Fourier transform of the magnetic field $b_i({\bf{x}})$
which is given in Alfven units so that $b_i = B_i /\sqrt{\mu_0 \rho}$. The solenoidal field fluctuations can be written in two parts as,   
\begin{equation}
  M_{ij}({\bf{k}}) = M^{dir}_{ij} ({\bf{k}}) + M^{pol}_{ij} {\bf{(k)}},
  \label{eq:Mdecomp}
\end{equation}
Here the directional part is the projection of $M_{ij}$ onto $P_{ij}$:
\begin{equation}
  M^{dir}_{ij}({\bf{k}}) = M^{dir}({\bf{k}}) P_{ij}({\bf{k}})
  \label{eq:Mdir}
\end{equation}
where $P_{ij}$ is the transverse projection matrix given by, $P_{ij} ({\bf{k}}) = \delta_{ij}- \frac{k_i k_j}{k^2}$. 
The second term on the right is the polarisation part given by, 
\begin{equation}
  M^{pol}_{ij} {\bf{(k)}} = M_{ij} {\bf(k)} - M^{dir}_{ij} {\bf{(k)}}.
  \label{eq:Mpol}
\end{equation}
For axial symmetry, it has been shown that the directional anisotropy is a scalar function of a wavevector argument.\cite{rubinstein} The $SO(3)$
decomposition for directional anisotropy is then reduced to a standard spherical harmonics
expansion. The polarisation anisotropy is the deviatoric part of the correlation
tensor and manifests itself directly as an imbalance in the energy contained in the fluctuations parallel to the axis versus those perpendicular to it. Also, the directional
anisotropy is restricted to even spins while the polarization anisotropy is manifested by
both the odd and the even spins with different operators. Since $M^{dir}{(\bf k)}$ is a scalar, its $SO(3)$ expansion is the standard
spherical harmonics series given by, 
\begin{equation}
  {M^{dir}{\bf (k)}
    = \sum_{\substack{l \geq 0 \\ l\;{even}}}
      \sum_{m=-l}^{l}
      A_{l,m}(k) k^{-l}Y^{l,m}(\bf{k})}.
  \label{eq:Mdir_expand}
\end{equation}
The scalar spherical harmonics $Y^{l,m}(\bf{k})$ are the standard homogeneous
harmonic polynomials. The coefficients are assumed to be symmetric under any interchange of 
indices. They are also traceless. For the axis symmetric case, the expansion simplifies significantly as there is only one polynomial for each spin. So the only label for
different modes is the index $l$.
\begin{equation}
  {M^{dir}{\bf (k)}
    = \sum_{\substack{l \geq 0 \\ l\;{even}}} 
      A_{l}(k) k^{-l}Y^{l}(\bf{k})}.
  \label{eq:Mdir_expand2}
\end{equation}
The coefficient $A_{l}(k)$ labelled by the index $l$ has the physical 
information about the different modes of the turbulent fluid. It has been shown previously \cite{bifarale} that each individual value of $l$ can have a different power law structure. 
The authors had used a perfectly homogeneous fluid flow and found out that  there is a hierarchical organization of the exponents as a function of $l$. They also found higher intermittency in the anisotropic structure compared to the isotropic structure. 

The polarization part has to be expanded appropriately, however this expansion requires the use of the differential operator formalism \cite{arad}. In this formalism, the required tensors can be 
obtained by operating on scalar functions with rotation-invariant matrices of differential 
operators given by $\mathcal{L}^l_{ij}$. This operator has the property that $\mathcal{L}^l_{ij} [\Phi_{l}(k)] $ is solenoidal for any $\Phi_l(k)$ which is homogeneous of degree $l$. Therefore, the expansion is given by,
\begin{equation}
  {M^{pol}{\bf (k)}
    = \sum_{\substack{l \geq 0 \\ l\;{even}}}
      \sum_{m=-l}^{l}
      B_{l,m}(k) k^{-l}Y^{l,m}(\bf{k})}.
  \label{eq:Mpol_expand}
\end{equation}
where the tensor spherical harmonics $Y^{l,m}_{ij} \bf(k)$ are defined by
\begin{equation}
  Y^{l,m}_{ij}{\bf(k)} + i\,Y^{l,-m}_{ij}{\bf(k)}
  = c_{l,m}\;\mathcal{L}^l_{ij}\ \left[
      (k_x+ik_y)^m\,\frac{\partial^m}{\partial k_z^m}\,Y^l \bf(k)
    \right],
  \label{eq:Ynumij}
\end{equation}
For the axis symmetric case only the degree $l$ is considered and $m = 0$, therefore we will have,  
\begin{equation}
  M^{pol}_{ij}{\bf(k)}
    = M^{pol} {\bf(k)}\;S^{pol}_{ij}{\bf(k)},
  \label{eq:Mpol_axi}
\end{equation}
where the axisymmetric polarisation tensor is given explicitly by
\begin{equation}
  {S^{pol}_{ij}{\bf(k)}
    = k^2 a_ia_j
      - ({\bf{a}}\cdot {\bf{k}})(k_ia_j+k_ja_i)
      + \frac{1}{2}({\bf a}\cdot{\bf k})^2\!\left[\delta_{ij}+k^{-2}k_ik_j\right]
      - \frac{1}{2} k^2\,P_{ij}\bf(k)},
  \label{eq:Spol}
\end{equation}
and the scalar $M^{pol}{\bf(k)}$ is expanded in the polarization Legendre functions
\begin{equation}
  M^{pol}{\bf(k)}
    = \sum_{l=0,2,4,\ldots} B_{l,0}(k) k^{-l}\ Z^l {\bf(k)},
  \label{eq:Mpol_Z}
\end{equation}
where
\begin{align}
  Z^0\bf(k) &= 1, \label{eq:Z0}\\
  Z^2\bf(k) &= 7({\bf a}\cdot{\bf k})^2 - k^2, \label{eq:Z2}\\
  Z^4\bf(k) &= 33({\bf a}\cdot{\bf k})^4 - 18k^2({\bf a}\cdot{\bf k})^2 + k^4. \label{eq:Z4}
\end{align}
Note that the $Z^l$ are different from the directional Legendre polynomials
$Y^l$. The directional and the polarisation anisotropy are orthogonal in function space. In 
all this discussions, we have not discussed the isotropic case as we are mainly interested in 
the anisotropic part.

This is the general method of measuring the directional and the polarization anisotropy in a 
turbulent plasma. As mentioned before, it is the integral values of $l$ in the multiscale correlation tensor which give us the different modes of the higher order fluctuations. Coherent structures are formed due to these higher modes of fluctuations. These 
are geometrical patterns generated both in velocity and magnetic fields. The most common structure 
being the vortex structure. In magnetic fields, the $l = 2$ order gives 
elongated magnetic fields referred to as the flux tubes while the 
$l = 4$ gives thin current sheets. These are the structures we are interested in. 

Plasmas with a strong magnetic field often are axis symmetric as the direction  of the field itself breaks the spatial isotropy. Though the 
field may be pointing in a certain direction, usually it is 
not a constant field. Fields generated through any of the different mechanisms proposed in the literature tend to stronger at certain points 
and decay as you move radially outward. The distribution of the field in the $r$, $\theta$ plane is most probably Gaussian. Such magnetic fields can be found around cosmic filaments \cite{carretti} and close to individual strings in a tangled cosmic string network. While Gaussian fields are common, it is not uncommon to have heavy tailed distributions
too \cite{sumandey}. In the next section, we study magnetic fluctuations based on the 
Gaussian and the Cauchy distributions to find the relevant lengthscales  for the coherent structures in these kind of magnetic fields.  

\section{The Gaussian distribution }
In this section, we probe magnetic field fluctuations given by a Gaussian distribution. As is well established, many inflationary magnetogenesis theories, generate the primordial
magnetic fields from quantum vacuum fluctuations. Since these quantum fields are typically free, their amplitude fluctuations  follow a Gaussian distribution. For such a distribution, the two point correlation function can be directly analysed using the equations discussed in the previous section. The Gaussian distribution is given by, 
\begin{equation}
{\bf B} = B_0 e^{-\rho^2/2R_0^2} \hat{z}
\end{equation}
Here $R_0$ gives the correlation length for the magnetic field fluctuations, $B_0$ is the amplitude. 
For a Gaussian function, the Fourier transform to the $k$ space remains a Gaussian. 
\begin{equation}
{ B}(k) = 2 \pi B_0 R_0^2 e^{- k^2 R_0^2/2}
\end{equation}
Here $k$ is the wave vector on the plane perpendicular to the magnetic field. 
We can then use eq.\ref{eq:Mdir_expand2} to obtain the correlation function 
for different values of $l$ (the angular momentum). Since only even values of $l$ are allowed and we are not interested in the isotropic case, the expansion will be given by, 
\begin{equation}
M^{dir}(\vec{k}) = A_2(k)  Y^2(\vec{k})k^{-2} + A_4(k)  Y^4(\vec{k}) k^{-4}
= C_2 k^{-2} + C_4 k^{-4}
\label{eq:ploteq}
\end{equation}
where the higher order terms are not considered as they do not give the coherent structures that we are interested in studying. The $l = 2$ sector is the
lowest contribution to anisotropy in homogeneous turbulence, the $m = 0$ (axisymmetric) terms are the terms we are interested in. For the $m = 0$ terms, 
we see that the anisotropy will also depend on the angle $\theta$, where $\theta$
is the polar angle between the wave vector and the axis of symmetry. Since our 
basis vectors are the spherical harmonics, we do the analysis in the spherical 
polar coordinates. Similar to the directional anisotropy, we also study the polarization anisotropy. In this case, we use the polarization tensor described
in eq.\ref{eq:Mpol_Z}. 
Similar to the directional case, here too only even values of $l$ are allowed so that we have, 
\begin{equation}
M^{pol}(\vec{k}) = B'_2(k) Z^2(\vec{k})k^{-2} + B'_4(k)  Z^4(\vec{k}) k^{-4}
= B_2 k^{-2} + B_4 k^{-4}
\label{eq:ploteq2}
\end{equation}

The tensor spherical harmonics so obtained gives the 
polarization anisotropy which is basically the difference between the poloidal and 
the toroidal fields. 

\subsection{Results}
\begin{figure}
\begin{subfigure}{0.5\textwidth}
	\includegraphics[width = \textwidth]{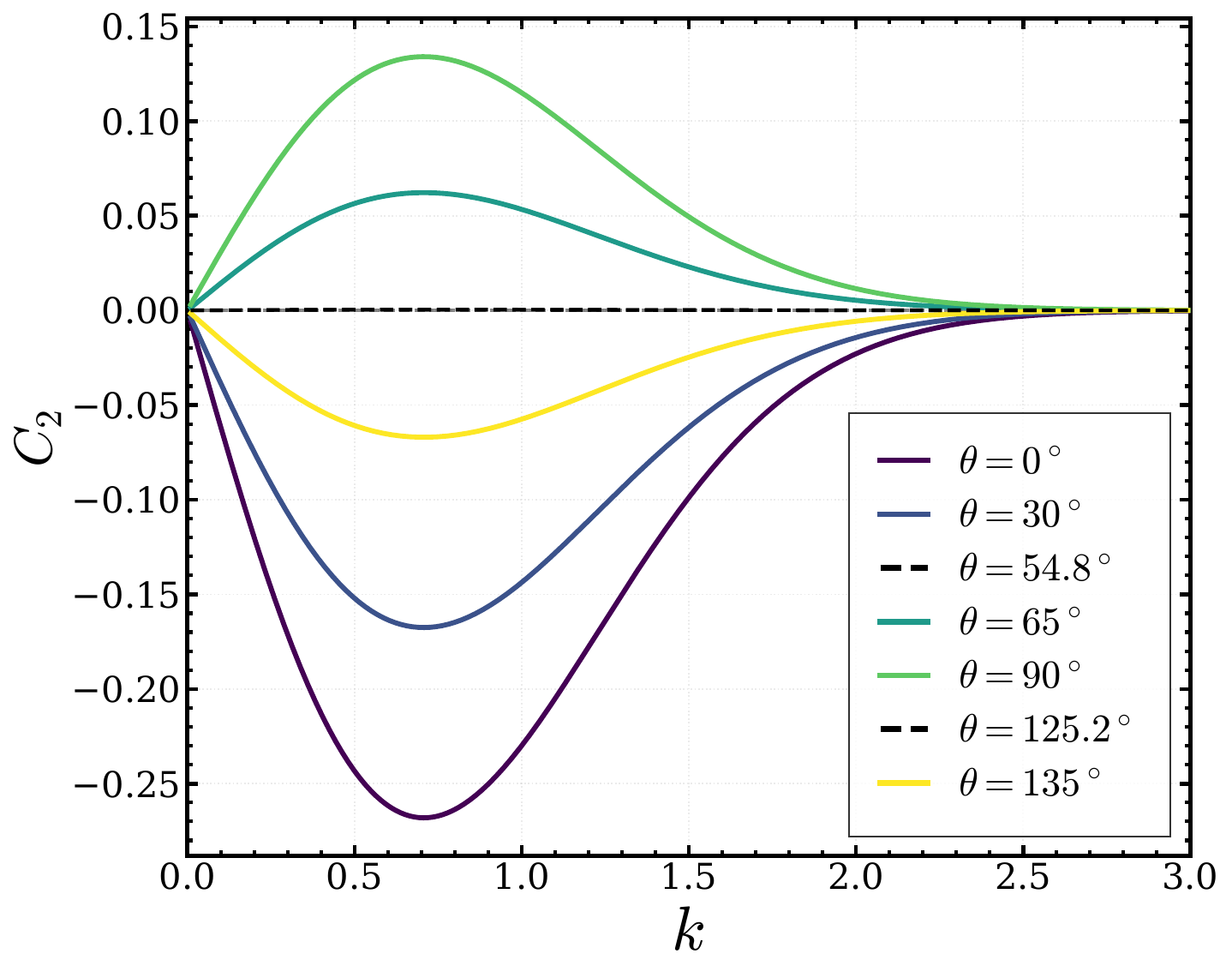}
	\caption{}
	\end{subfigure}%
\begin{subfigure}{0.5\textwidth}
	\includegraphics[width = \textwidth]{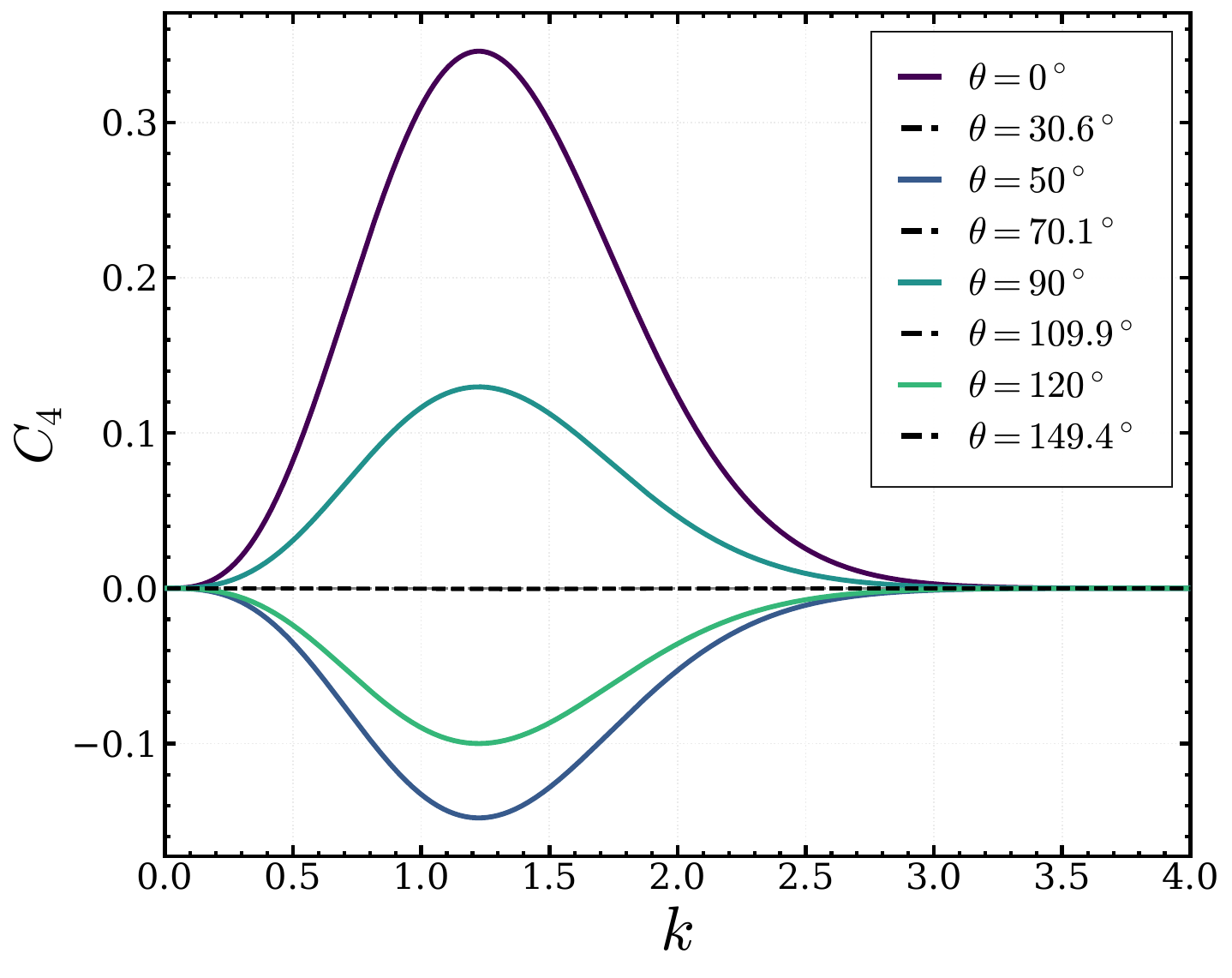}	
	\caption{}	
	\end{subfigure}%
\caption{The graph on the left has the $ l= 2$ mode for directional anisotropy for different values of the angle $\theta$ for a magnetic field with a Gaussian distribution. On the right, we have the $l = 4$ mode for the same field.}

	\label{fig:Gaussdir}
\end{figure}

Fig.\ref{fig:Gaussdir} shows the coefficient ($C_2$) of the magnetic field correlation  for the $l=2$ 
term for directional anisotropy.The anisotropy is scale dependent, and it also depends on the polar angle $\theta$. 
The correlation length is taken to be $R_0 = 1$ and the peak of the 
anisotropy occurs at $k \sim 0.7$. The value of $k$ for which the anisotropy reaches a maximum 
is approximately given by $k \sim \frac{1}{\sqrt{2} R_0}$. We have changed the correlation length 
and found that the peak shifts accordingly.    
In fig.\ref{fig:Gaussdir}, we have also
plotted the $l = 4$ mode. The $\theta = 0$ or polar angle is positively correlated 
for the $l = 4$ mode. The correlation here is stronger than the $l = 2$ mode. For 
the $\theta = \frac{\pi}{2}$ however the $l = 2$ mode is stronger. As mentioned 
before, the $l = 2 $ mode corresponds to the flux tubes while the $l = 4$ mode
is for the thin current sheets. The toroidal component then gives elongated flux 
tubes while the poloidal component are stretched into sheet like structures. 
This kind of poloidal toroidal twisted shape has been observed in numerical simulations of astrophysical objects \cite{braithewaite}.

In both the cases we have considered $B_0 = 1$. Increasing the amplitude just increases the 
anisotropies proportionately hence we have kept  $B_0 = 1$ throughout this work. 
We see that the sign of the correlation gets flipped periodically indicating that the magnetic 
correlation in the coherent structures for $l = 2$ will have different rotational 
properties relative to the magnetic field. The sign reversal is due to the fact that we have considered
homogeneous MHD turbulence which can be represented using the $SO(3)$ representation. 
We will also see this kind of feature appearing for the $l = 4$ term. 
The sign reversal in the coherent structure  occurs due to the geometry of the spherical harmonics with the reversal happening at a predetermined angle $\theta$. These are 
also referred to as the {\it magic angles} in rotational turbulence. At these particular 
angles the magnetic correlation is zero. This reversal of the sign results 
in the acceleration of the particles at certain angles of $\theta$.

\begin{figure}
\begin{subfigure}{0.5\textwidth}
	\includegraphics[width = \linewidth]{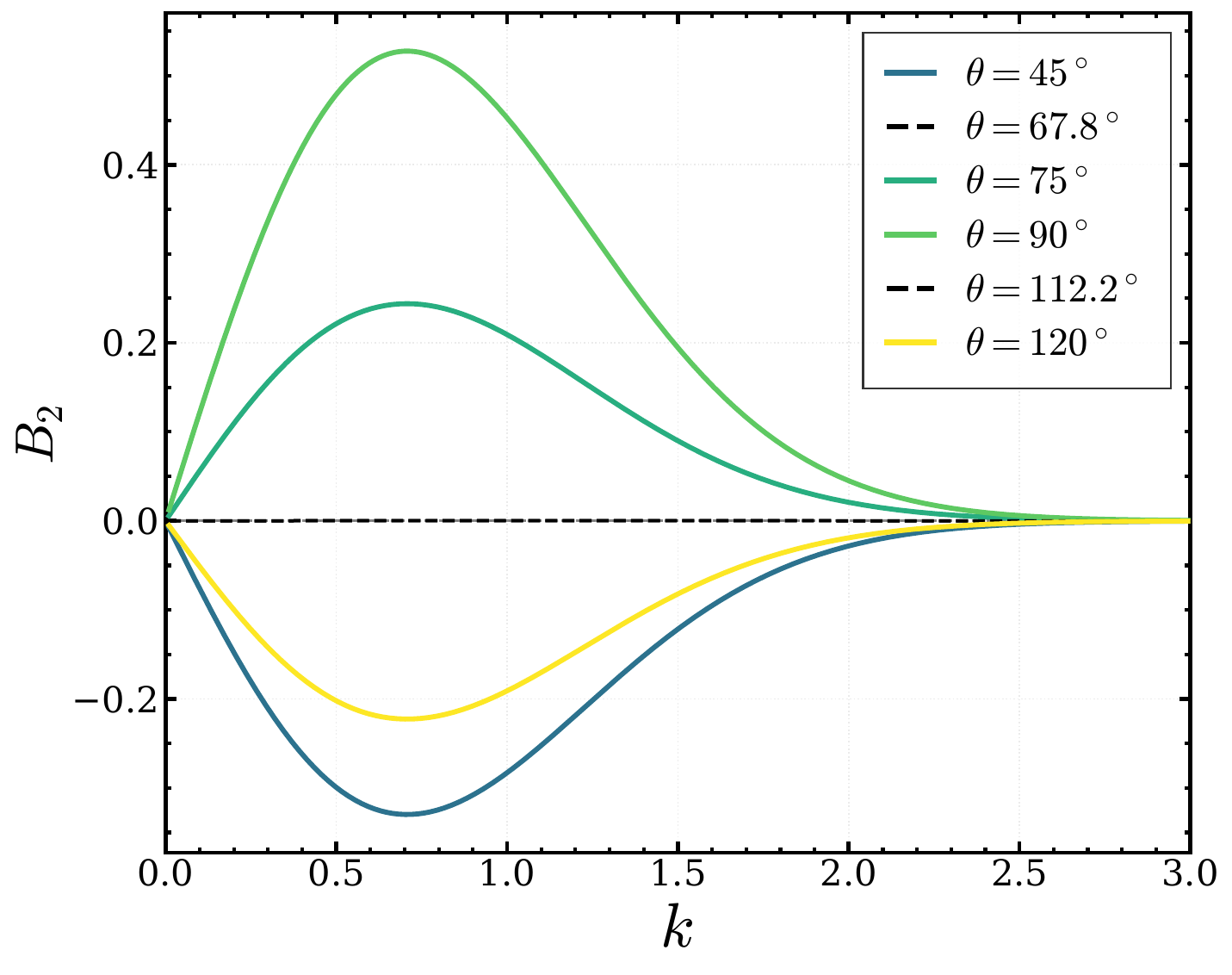}
	\caption{}
	\end{subfigure}%
\begin{subfigure}{.5\textwidth}
	\includegraphics[width = \linewidth]{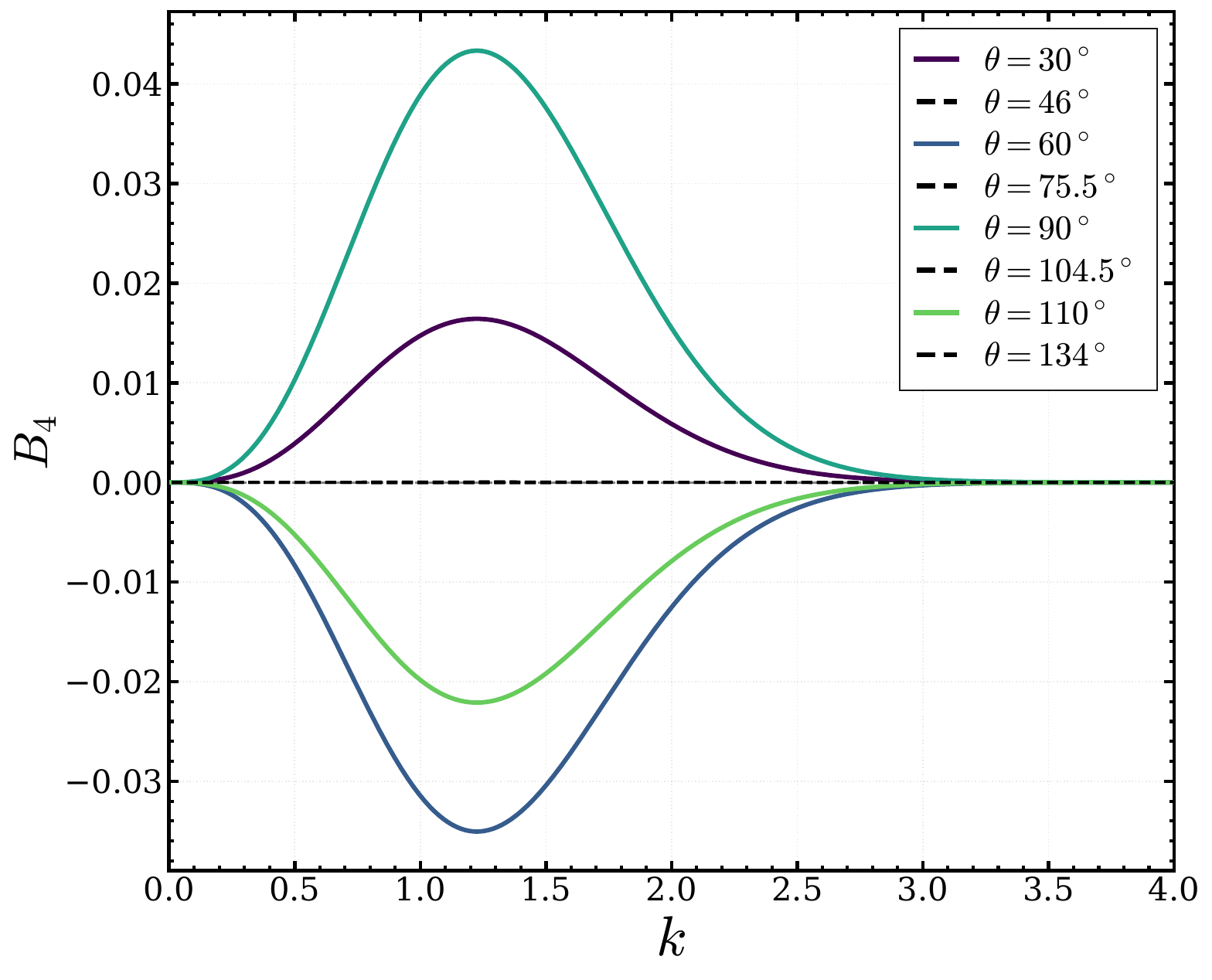}
\caption{}
	\end{subfigure}%
	\caption{The $l = 2$ and the $l = 4$ mode for polarization anisotropy for different values of the angle $\theta$ for a magnetic field with a Gaussian distribution }
		
	\label{fig:Gausspol}
\end{figure}

 We now look at the polarization anisotropy in fig \ref{fig:Gausspol}. Overall, the distribution of directional anisotropy in terms of degree and wave number looks similar to its corresponding one for polarization anisotropy. However, 
in this case, the $l=2$ mode anisotropy clearly dominates over the $l = 4$ mode. 
The overall asymmetry in the positive and negative correlation is also smaller. Interestingly, the 
polarization anisotropy for $l = 2$ is greater than the directional anisotropy in magnitude for some 
angles. When polarization anisotropy dominates over directional stretching, the field fluctuations are fixed into certain polarization states. This will lead to the slowing down of the energy transfer from large to small lengthscales.

\section{The Cauchy distribution }
Though primordial magnetic field fluctuations generated from inflationary sources and several other sources 
are Gaussian in nature, there is the possibility of having non-Gaussian fluctuations too.  Generally, these are modelled by some heavy tailed distribution. Here we choose the Cauchy distribution to 
probe the anisotropic structures in a non Gaussian magnetic field. The heavy tail of the Cauchy distribution allows for the persistence of the magnetic fluctuations at larger lengthscales and therefore allows us to probe the smaller $k$ region. The  distribution is given by, 
\begin{equation}
{\bf B} = \frac{B_0}{1 + \left(\frac{\rho}{\gamma}\right)^2} \hat{z}
\end{equation}
where $\gamma$ is the full width at half maxima for the magnetic fluctuation. 
The magnetic fluctuation is defined in the cylindrical coordinate system. 
This is useful in obtaining the 
correlation function in the spectral space as the Fourier transform of a multivariate Cauchy distribution can be expressed as a modified Bessel function. In the wave vector space therefore we have, 
\begin{equation}
B(k) = 2 \pi B_0 \gamma^2 [K_0(\gamma k)]
\end{equation}
where $K_0$ is the modified Bessel function of the second kind of order 
zero. We again look only at the $l = 2$ and $l = 4$ components of eq.\ref{eq:ploteq}.

\subsection{Results}
\begin{figure}[t]
\begin{subfigure}{0.5\textwidth}
	\includegraphics[width = \linewidth]{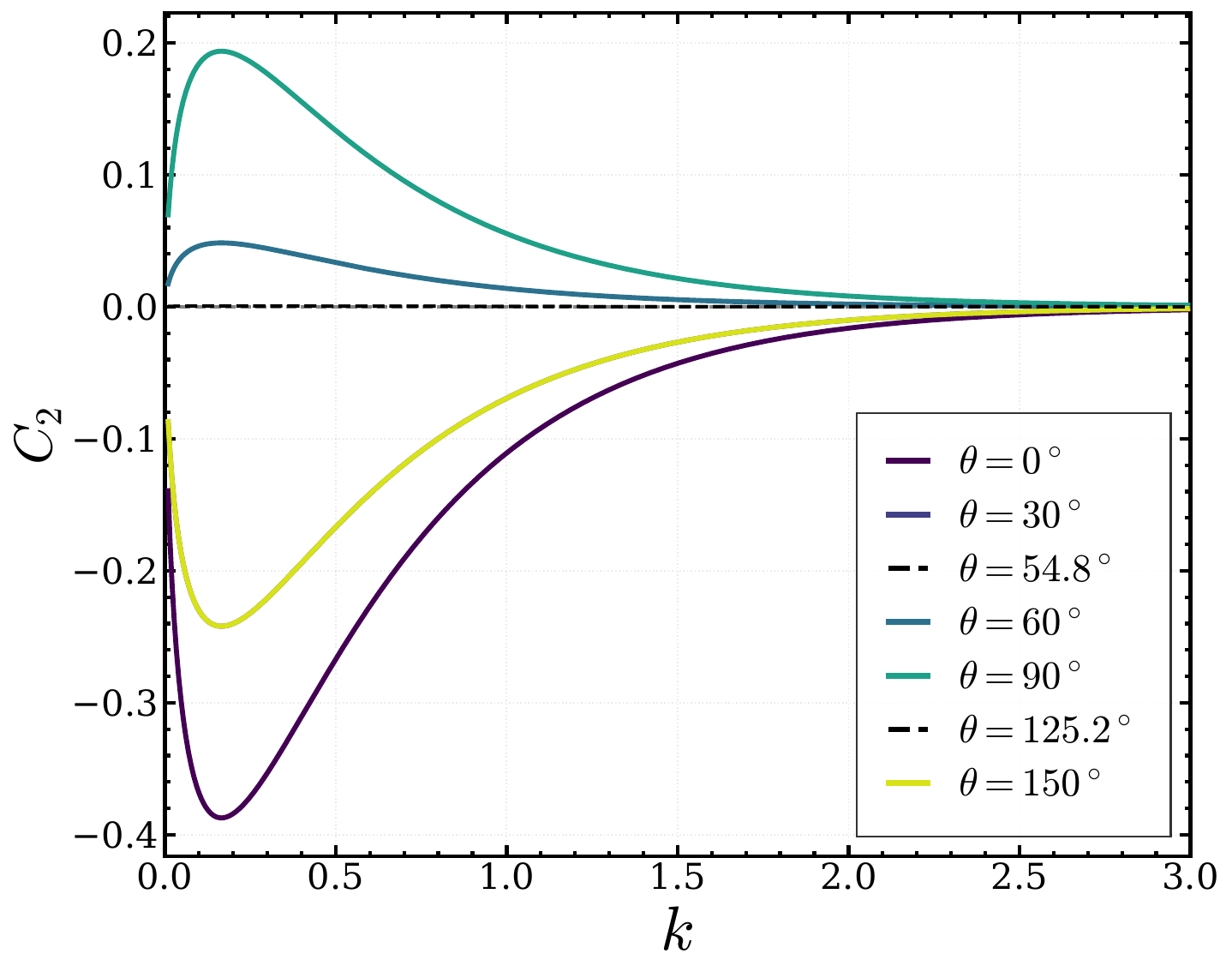}
	\caption{}
	\end{subfigure}%
\begin{subfigure}{.5\textwidth}
	\includegraphics[width = \linewidth]{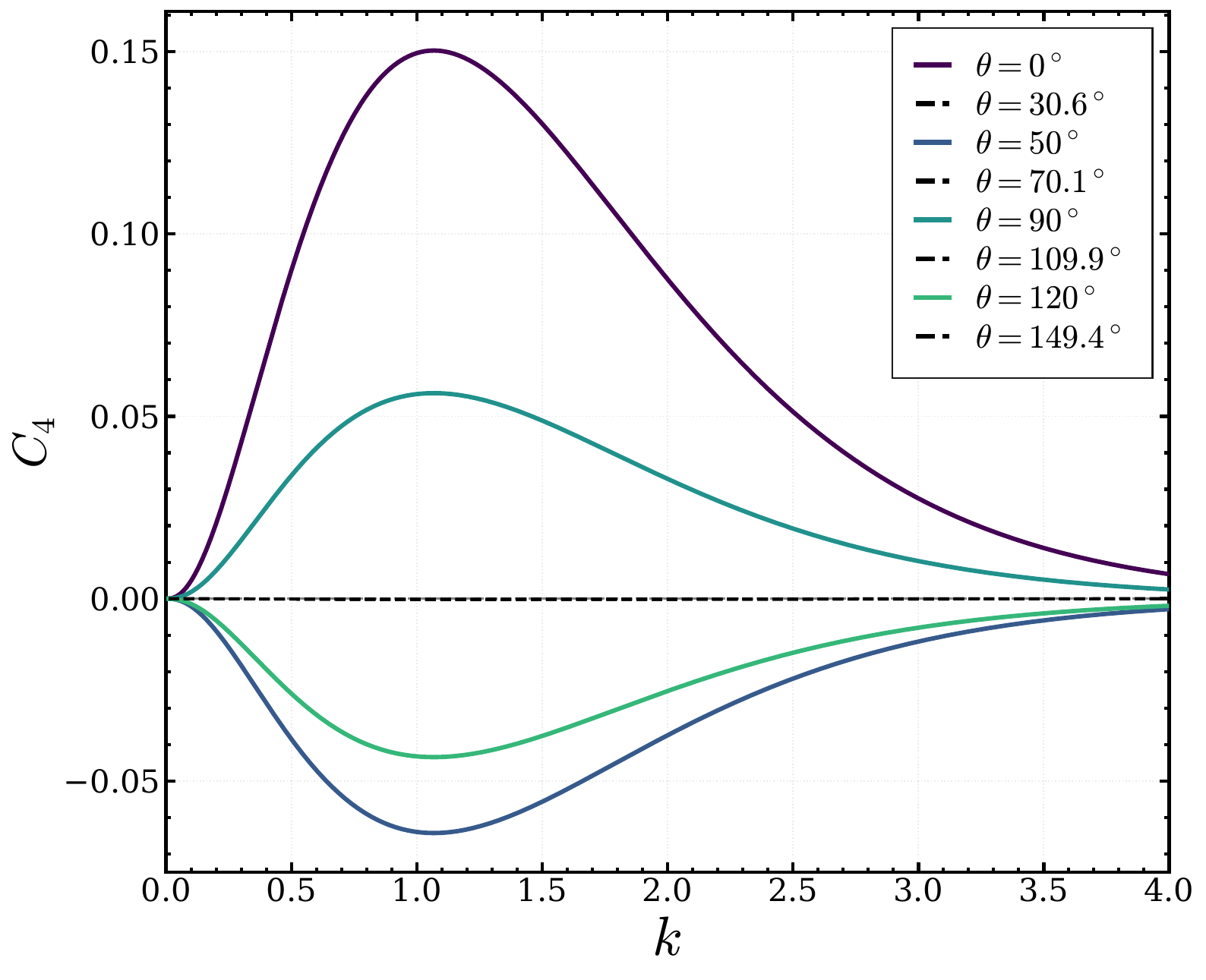}
	\caption{}
	\end{subfigure}%
	\caption{The $l = 2 $ and the $l = 4$ mode for directional anisotropy for different values of the angle $\theta$ for a magnetic field with a Cauchy distribution}
		\label{fig:Cauchydir}
\end{figure}

\begin{figure}
\begin{subfigure}{0.5\textwidth}
	\includegraphics[width = \linewidth]{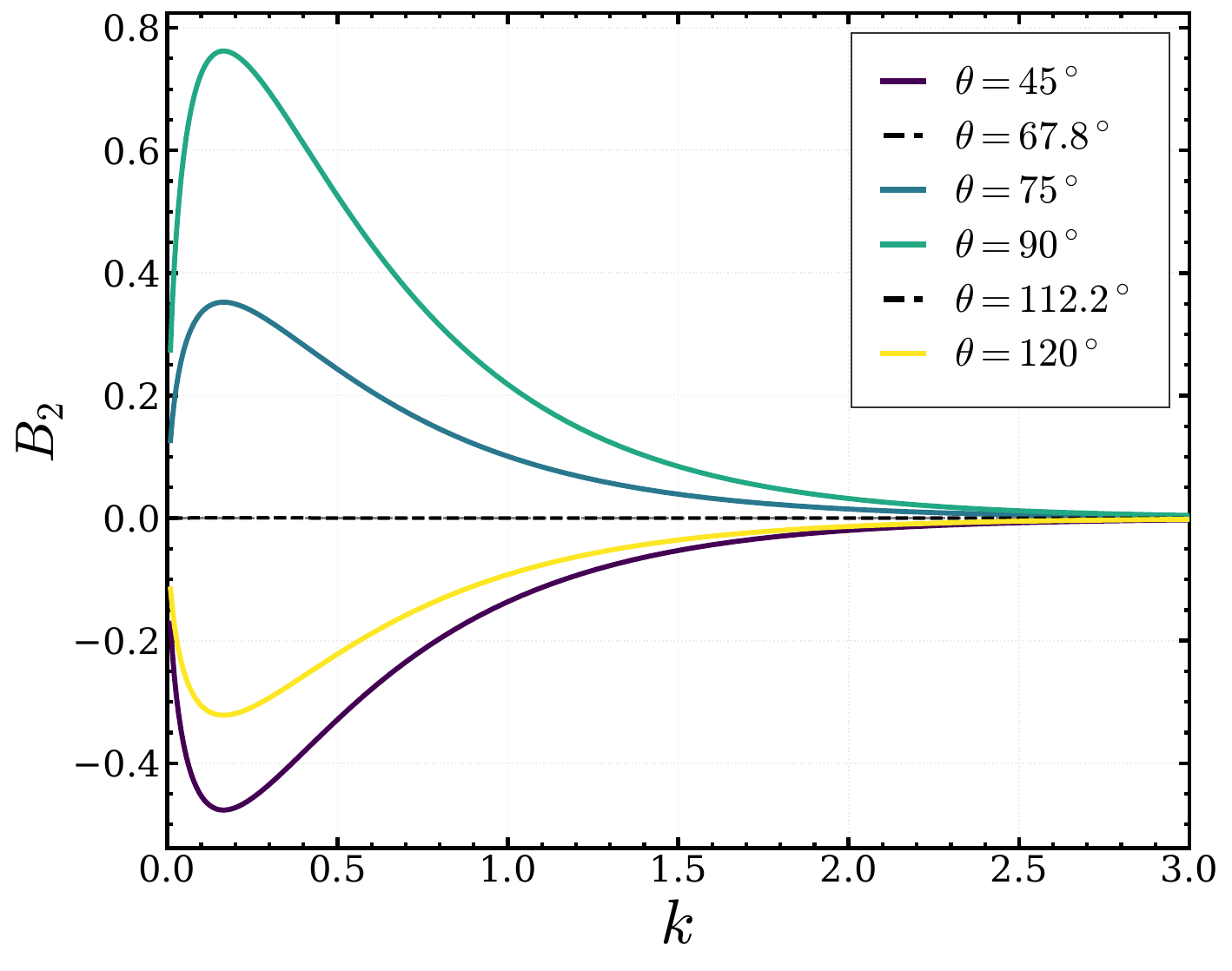}
	\caption{}
	\end{subfigure}%
\begin{subfigure}{.5\textwidth}
	\includegraphics[width = \linewidth]{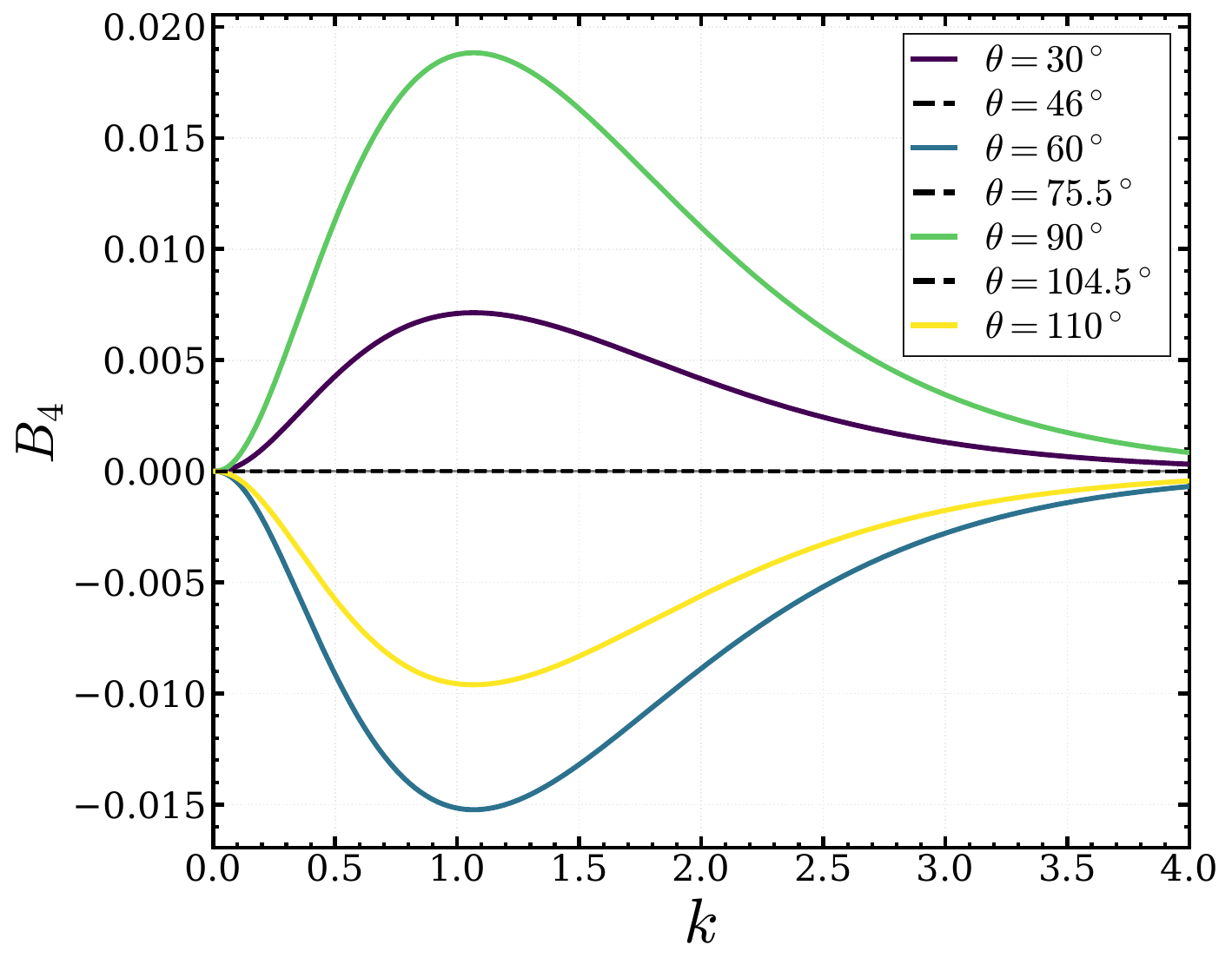}
\caption{}

	\end{subfigure}%
		\caption{The $l = 2 $ and the $l = 4$ mode for polarization anisotropy for different values of the angle $\theta$ for a magnetic field with a Cauchy distribution }
		
			\label{fig:Cauchypol}
\end{figure}

For the graphs, we have considered the parameters $\gamma = 1$ and $B_0=1$. 
Fig.\ref{fig:Cauchydir} shows the correlation in the magnetic field for the $l=2$ and the $l=4$ 
term for the directional anisotropy. The basic features of the correlation spectrum remains the 
same except that the peak of the anisotropy now occurs close to $k=0$ which is expected for the 
Cauchy distribution. In this case, changing the correlation length changes the way the field decays at higher values of $k$. The flipping of the positive and negative correlation occurs here too, due to the same reasons as discussed previously. The anisotropies are higher at smaller $k$ values in this case.      
In the $l = 4$ mode, the anisotropies are of similar order as the $l = 2$ mode for the positive correlation whereas they are about an order of magnitude smaller than the second order anisotropies when 
the fluctuations are negatively correlated. While most of the conclusions are therefore unchanged, 
the Cauchy distribution displays a steeper gradient and a longer tail.    
The polarization anisotropy for both the modes are given in fig.\ref{fig:Cauchypol}. The steeper gradient 
would mean that the polarization states in this case would lead to sharper changes in certain lengthscales. The flipping of the correlations and the longer range of lengthscales involved would mean 
that the motion of the particles would crucially depend on their diffusion length. Similar to the 
Gaussian distribution, the field gradients here are also forced into rigid spatial configurations.
Only in this case, the field gradients at longer lengthscales (smaller $k$) are steeper. We also find that 
the correlation remains non zero for a longer range of $k$ for the modes, especially for the $l = 4$ mode. This 
indicates that the anisotropic fluctuations of the magnetic field persist for much smaller lengthscales compared to the Gaussian distribution.  

All these show that the coherent magnetic structures have a form which is well defined and though the 
Gaussian and Cauchy distributions apparently look similar their structure is different.  The Cauchy 
distribution leads to multiple lengthscales and steeper gradients compared to the Gaussian distribution.

Usually a plasma scatters particles randomly. This is modeled by a stochastic process
where particles diffuse through random, small-scale fluctuations of the magnetic field.
In the case of coherent structures though, the particles are scattered in a deterministic path  which involve localized and ordered magnetic geometries. A positive
correlation function indicates that the magnetic fields at the two points are aligned to
each other whereas a negative correlation function means that over the short distance, 
the two fields are antiparallel to each other. Between these two, there is a short line 
where the correlation is zero. This immediately indicates that the particle motion is 
subjected to sudden changes of path. As has been seen from previous
studies, the two-point magnetic correlation function for homogeneous, axisymmetric turbulence affects the mean free path of the charged particles \cite{shalchiArendt}. We will show that the diffusion coefficient will also depend on the distribution of the magnetic field fluctuations in the 
next section.   

\section{Astrophysical Implications}
Magnetohydrodynamic turbulence is present at different epochs in the early universe. 
Coherent structures are an integral part of magnetic turbulence. While it has been studied extensively for solar and intergalactic physics \cite{Ntourmousi}, it is less discussed in the context of the early universe. It is however postulated that magnetic turbulence and the dynamo effect are responsible for the growth and evolution of the 
seed magnetic field produced in the early universe. The early universe has a plethora of axisymmetric magnetic fields. Current carrying cosmic strings \cite{krtous,zadorozhna} as well
as magnetic fields generated around cosmic strings via the Biermann mechanism \cite{sovan,bisht} all result in axis - symmetric magnetic fields. MHD turbulence will 
lead to coherent structures in these axisymmetric fields. Generally, a Gaussian 
distribution is the accepted field distribution for these kind of field, however as we 
have seen in our analysis, it is important to understand the underlying magnetic field 
distribution as a heavy tailed distribution and a Gaussian distribution will give different results for the scattering of particles from these 
coherent structures. It is only recently that the role of coherent magnetic 
structures and the theories of geometry mediated transport are becoming 
important in understanding the transport of high energy cosmic rays \cite{lubke}. 

The scattering of high energy charged particles in the astrophysical plasma depends on the 
magnetic field correlation vector. Particle scattering is anisotropic even in a magnetized plasma where only isotropic fluctuations are considered \cite{shalchi}. 
The diffusion coefficient of a particle in a magnetized plasma is given by, 
\begin{equation}
\kappa_{ij} = \int_0^{\infty} dt <V_i(t) V_j(0) > 
\end{equation}
where $<V_i(t) V_j(0) > = V_{ij} $ is the velocity correlation function. The velocity correlation function can be expanded to include the magnetic field fluctuations at the position of the particle. In case the velocity correlation of the particle is uncorrelated to the magnetic field 
correlation we can simplify the fourth order correlation as a product of the two point velocity correlation and the two point magnetic field 
correlation. This can be done by assuming that the movement of the magnetic field line is the most important mechanism leading to particle scattering in the direction perpendicular to the mean magnetic field. We can then obtain the diffusion coefficient in the perpendicular direction as, \cite{shalchi}
 \begin{equation}
\kappa_{ij} = \int_0^{\infty} dt \frac{M_{ij}({\bf r}(t))}{B_0^2 + M_{zz}({\bf r}(t))} V_{zz}(t)
\label{eqn:kappa}
\end{equation}
Here $V_{zz}(t)$ is the velocity correlation function along the mean magnetic field. $M_{ij}({\bf r}(t))$ is the magnetic field correlation in the physical space. Since we 
have not considered the velocity correlation in this work, for the time being we will consider
$V_{zz}(t)$ to be a constant given by it's maximum possible value $V_{zz}(0)$ (it's zero lag value). We shall make use of Corrsin’s independence hypothesis \cite{tautz} and consider that 
the correlation depends solely on the spatial difference $|r - r'|$, therefore we are interested in,  
\begin{equation}
M_{ij}({\bf r}(t)) = \int d^3r  R_{ij}({\bf{r}}) f_p({\bf{r}};t)
\end{equation}
where $f_p({\bf{r}};t)$ is the particle distribution in the plasma at a given time. 
Here $R_{ij} =  <B_i({\bf r})B_j({\bf 0})>$ is the magnetic field correlation in the physical space.In the presence of higher order fluctuations, the particle scattering is anisotropic 
but in a coherent predictable way. The diffusion coefficient of the particles would also depend on particle distribution function. Assuming the particle distribution function to be constant 
in time,  we can also obtain an upper bound on the 
diffusion coefficient for the two distributions using the Cauchy-Schwarz inequality
\cite{papatha,kheruntsyan}. The Cauchy Schwarz inequality as applied to fluctuating quantities states that the expectation value of the cross correlation between any two quantities is bounded
from above by the autocorrelation in each quantity. In eq.\ref{eqn:kappa}, we find that the 
diffusion coefficient is a product of the fluctuations of the magnetic fields and the velocity 
fields. Here we have considered the plasma to be in local thermal equilibrium and therefore we
can consider the particle distribution (i.e the density) to be homogeneous and independent of 
time. So the plasma does not contain any density fluctuations.  
As mentioned previously, the autocorrelation for the velocity field is bound by $V_{zz}(0)$ for both the magnetic field distributions. Then we will have 
\begin{equation}
|\kappa_{ij}| \leq [R_{ij}] [V_{zz}(0)]
\end{equation} 
where $[R_{ij}]$ is the autocorrelation for the magnetic field distributions we have considered.   
We have obtained the correlation  of the field distributions in the wave vector 
space. The autocorrelation of the Gaussian remains a Gaussian with a wider width, while for the
Cauchy distribution the autocorrelation will depend on the cut-off lengthscale of the distribution.  So while for the Gaussian field distribution, the diffusion coefficient due to the magnetic field decays exponentially, for the Cauchy distribution it's decay can 
be approximated as the square of the Bessel function of the second kind depending on how the
distribution is truncated. We also see that the constants $C_2$, $C_4$, $B_2$,$B_4$ become zero at certain values of $\theta$, this indicates that at these angles the fields become completely 
uncorrelated on this plane. These angles however remain the same for both the
distributions. So it is the nature of the distribution which is crucial for the scattering 
of the particles from these coherent structures.  

This also leads to an understanding of the different scales of diffusion in these coherent structures. As the Gaussian decays exponentially, there is a $k_{max}$ beyond which the correlation is zero for all the curves. It means that in physical space, the fields are perfectly smooth below a minimum lengthscale $\lambda_{{min}} = 2\pi / k_{max}$. This describes
the Kolmogorov microscale for the smallest eddies. It is a finite localized scale where all the fluctuations are suppressed. In the Cauchy distribution though we find that the correlation goes to zero at much larger values of $k$ compared to the Gaussian distribution. Therefore, 
the angular spectra for the non Gaussian distribution is over a larger lengthscale going upto even very small length scales.

At small values of $k$, again the Gaussian goes smoothly to zero, while the squared modified Bessel function has a logarithmic divergence close to zero. 
This indicates a long range connectivity for the non-Gaussian distribution. For the mid range of $k$ values we also find a drop in the correlation values which has a different slope compared to the very smal range vaues of $k$. This would give us an intermediate lengthscale for this distribution.

\begin{figure}
\begin{subfigure}{0.5\textwidth}
	\includegraphics[width = \linewidth]{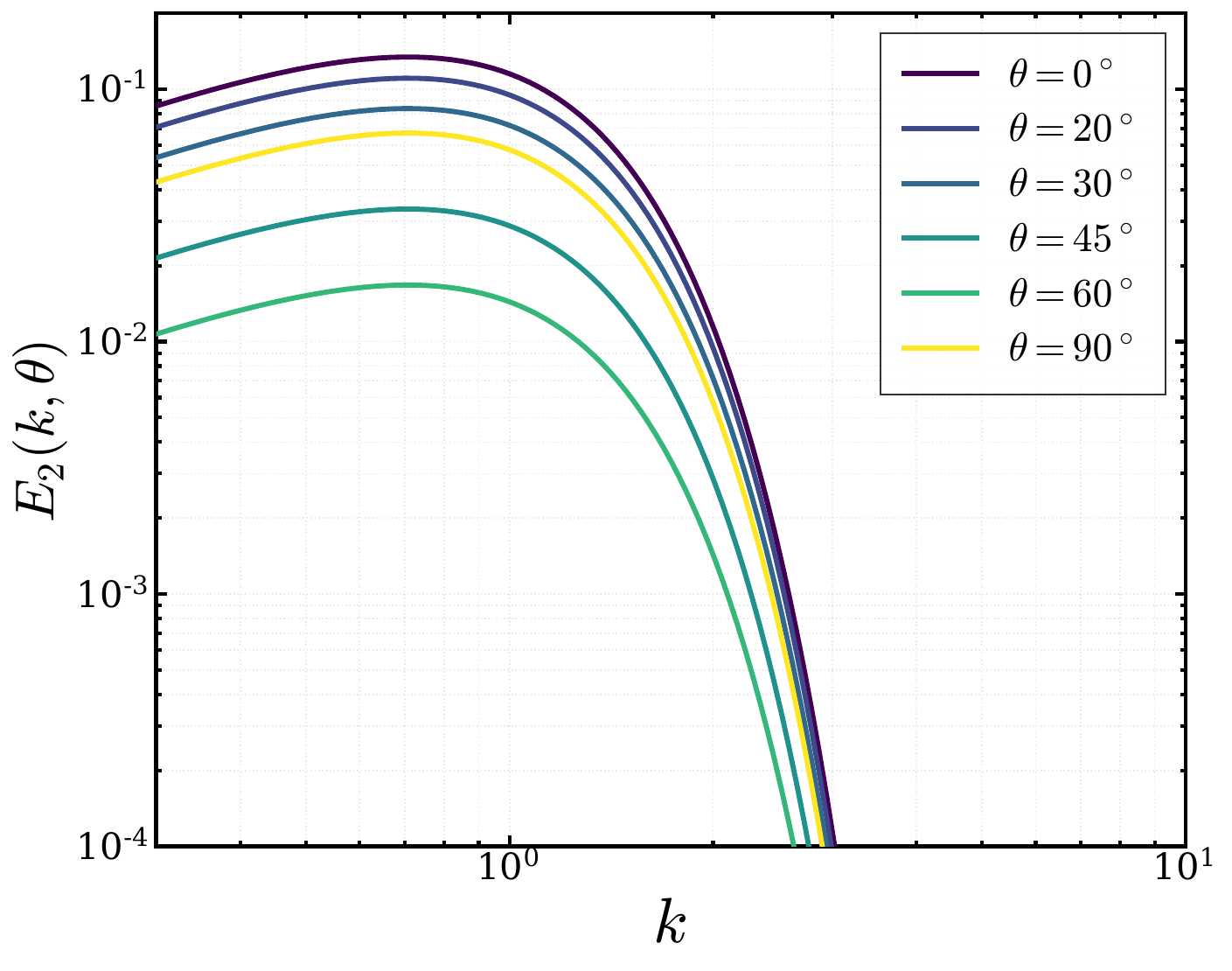}
	\caption{ The Gaussian distribution.  }
	\label{fig:Gaussenergyspectra2}
	\end{subfigure}%
\begin{subfigure}{.5\textwidth}
	\includegraphics[width = \linewidth]{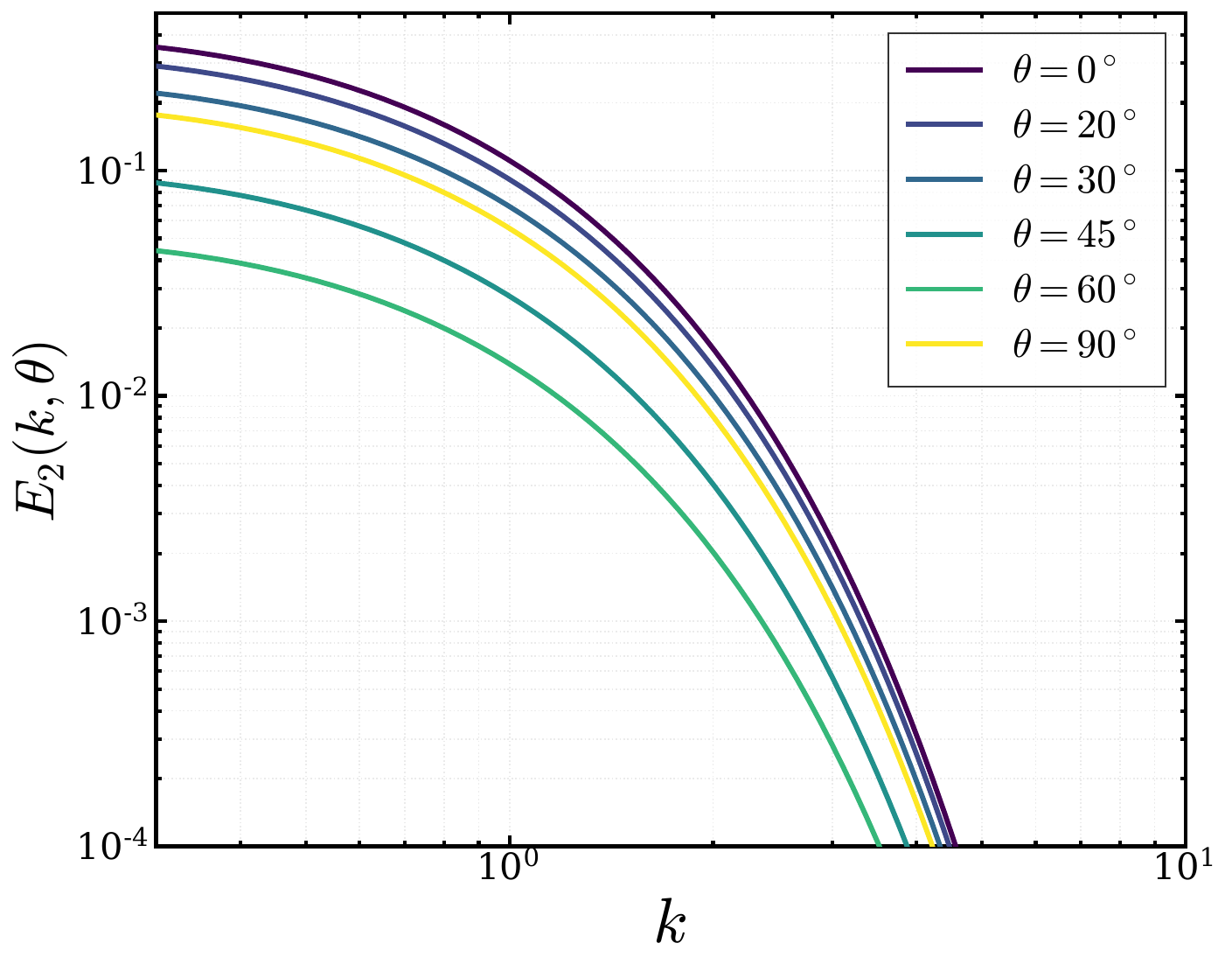}
	\caption{ The Cauchy distribution.}
	\label{fig:Cauchyenergyspectra4}
	\end{subfigure}%
	\caption{The Energy Spectrum for $l = 2$}
	\label{fig:energyspec}
\end{figure}

\begin{figure}
\begin{subfigure}{0.5\textwidth}
	\includegraphics[width = \linewidth]{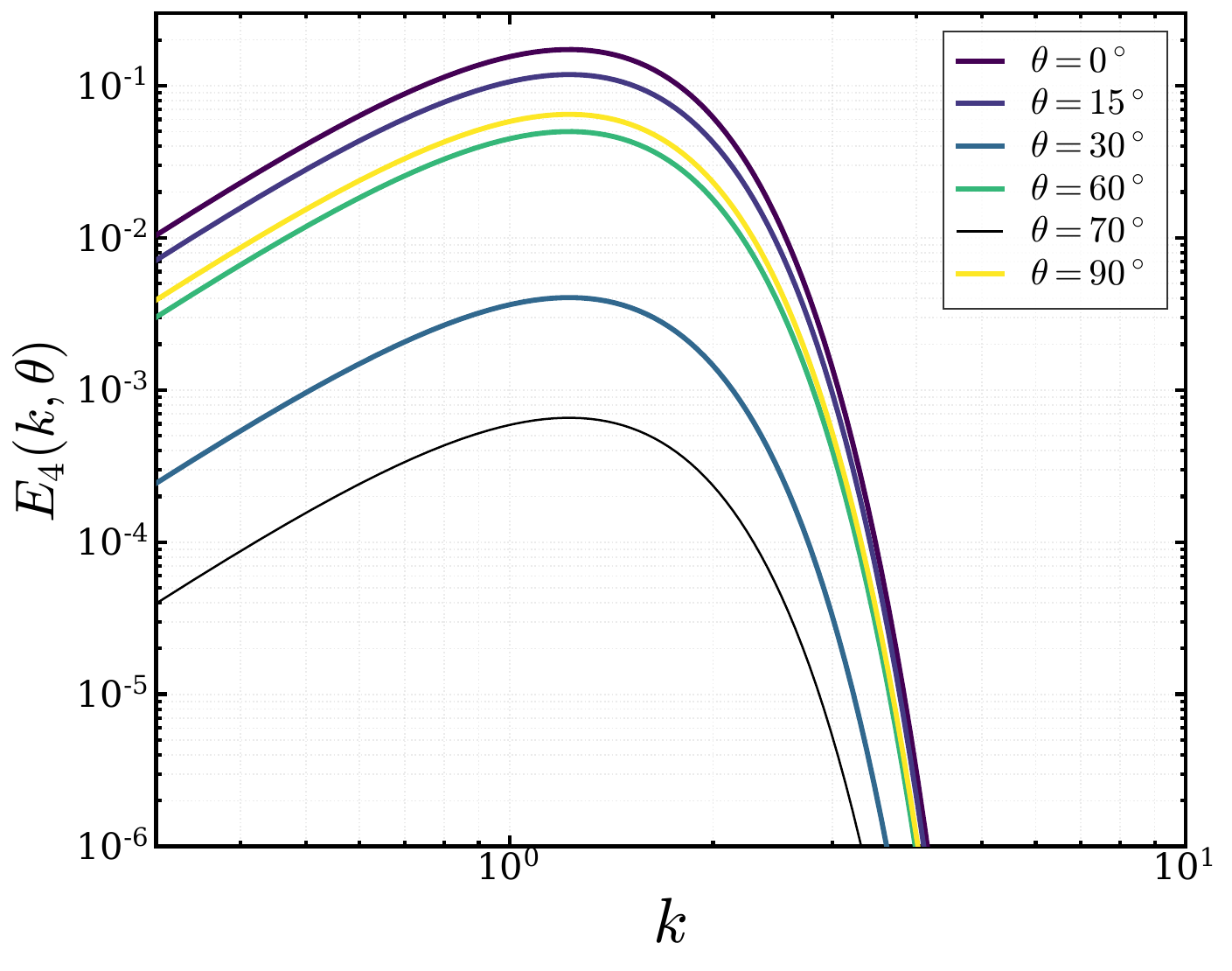}
	\caption{ The Gaussian distribution.  }
	\label{fig:Gaussenergyspectra4}
	\end{subfigure}%
\begin{subfigure}{.5\textwidth}
	\includegraphics[width = \linewidth]{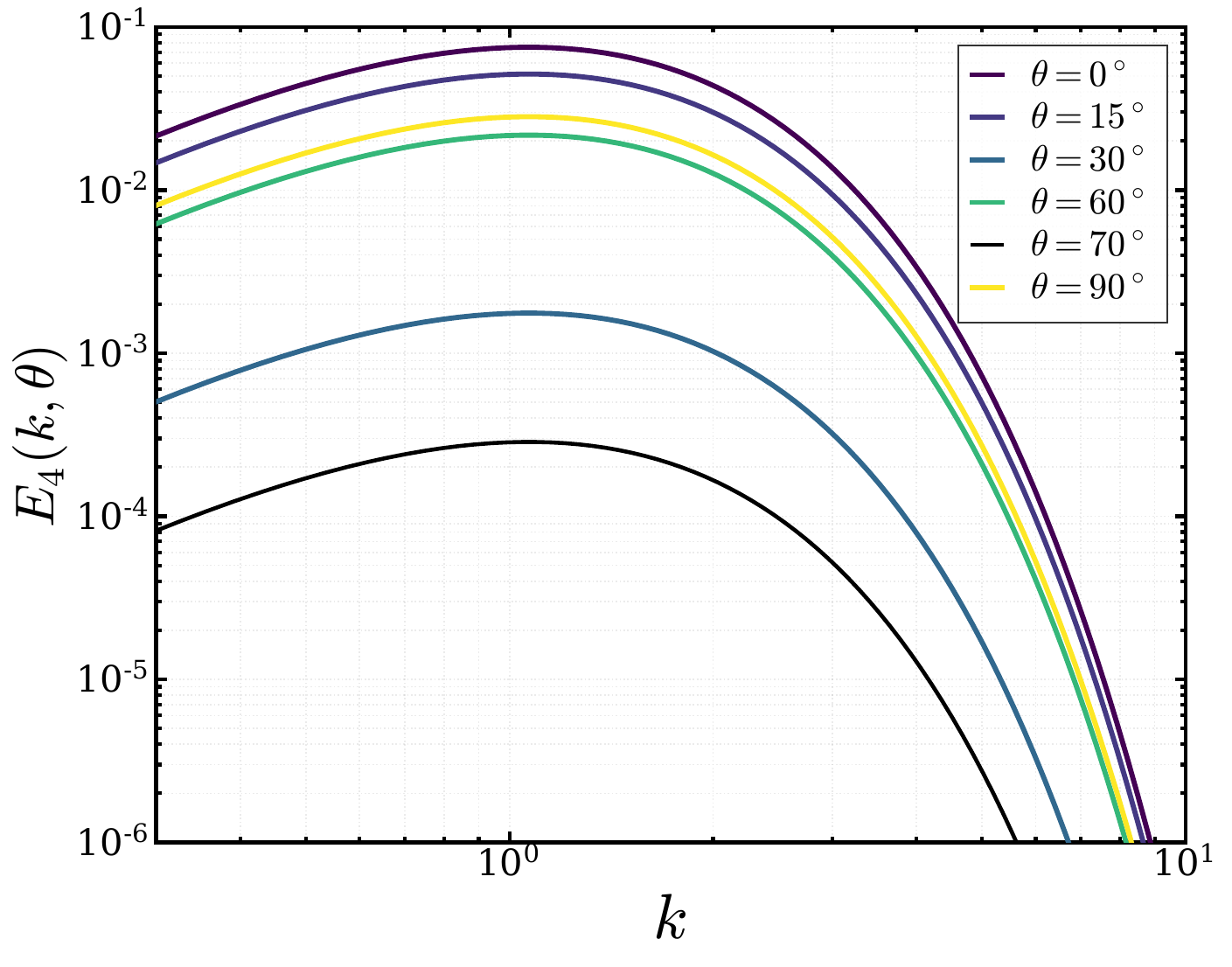}
	\caption{ The Cauchy distribution.}
	\label{fig:Cauchyenergyspectra2}
	\end{subfigure}%
	\caption{The Energy Spectrum for $l = 4$}
	\label{fig:energyspec2}
\end{figure}

The magnetic energy spectra for the Gaussian homogeneous magnetohydrdynamic turbulence for 
different angles of $\theta$ for $l = 2$ is shown in fig. \ref{fig:Gaussenergyspectra2}. The energy spectrum has a broad inertial range 
followed by a steep drop, so only two lengthscales can be identified from here. The magnetic energy spectra at different angles of $\theta$ ($l=2$) for the Cauchy distribution is 
given in fig.\ref{fig:Cauchyenergyspectra2}.  We also obtained the energy spectrum for the $l = 4$ mode. This is shown in fig. \ref{fig:energyspec2}.

The coherent structure correponding to the Gaussian distribution  has a smoother gradient between the positively and negatively correlated regions. There are essentially two lengthscales involved, the integral coherence length, 
which measures the spatial extent of the largest coherent eddies and the Kolmogorov microscale. Given our distribution, 
the integral coherence length is given by, $k_1 \sim \frac{1}{\lambda_c} $ where $\lambda_c \approx \frac{\pi}{4} $.

The energy spectrum of the Cauchy spectrum has a different slope as it decays. Since the decay is not so abrupt, one can define an intermediate lengthscale for the magnetic correlation apart from the two in the Gaussian distribution.  So essentially for the Cauchy distribution we can have three zones. Typically  when there is an intermediate range between the smallest and the largest eddies, one can use the Taylor microscale as the lengthscale to characterize the turbulent flow in this 
region. The standard Taylor microscale is normally defined from the curvature of the correlation function at zero separation. Since this is a local quantity, it only needs the correlation function to be smooth at $r = 0$. So even though the global variance is undefined for a Cauchy 
distribution, one can still define a local Taylor microscale for this distribution.

Physically also, a magnetic field does not extend to 
infinity, it always has a finite cut-off. This finite cut-off allows us to define the lengthscales for such a distribution. The coherence lengthscale (corresponding to $k_1 $) will be given by approximate values such as the full width at half maxima $(\gamma)$ of the distribution or a truncated Cauchy 
lengthscale related to $(\gamma)$.  The Taylor microscale can be obtained based on the dissipation rate and the root mean square velocity of the particles.  This can be related to the wave vector $k_2$. Between $k_1$ and $k_2$ the physical behaviour changes significantly and there are sharp deflections. Beyond $k_2$, the energy dissipates very fast, this is in the region of the Kolmogorov microscale. The field is tangled and the particles lose energy. The angle $\theta$ also plays an important role in 
whether the particles are moving in a coherent way or in a chaotic manner. This is because 
the correlation strength and direction depend upon this $\theta$. Thus the diffusion of particles are
modified considerably by the coherent structures.

The diffusion of particles in the early universe is important for the growth and decay of density inhomogeneties \cite{das}. Density inhomogeneities at temperatures 
below the Quark - Hadron transition lead to inhomogeneous nucleosynthesis scenarios. This 
is especially true for charged density inhomogeneities which will affect the path of the 
proton \cite{abhijit}. In the presence of a magnetic field, this is further
affected by the Lorentz force acting on the particles. For neutral particles such as the the neutron, the scattering cross section will not be affected directly by the magnetic field correlations. So the path of the proton and the neutron as well as their drift velocities will be significantly different in the coherent structures. This ultimately affects the neutron proton ratio in localized regions giving rise to inhomogeneous nucleosynthesis.

Usually in most cases, the primordial magnetic field fluctuation is taken to have a Gaussian distribution, however it is important that we consider non - Gaussian distributions too, especially the heavy tailed distributions.  For a heavy tailed distribution, we find three distinct lengthscales 
in which the diffusion coefficients are different. While the particles will not be deflected 
significantly around the coherence lengthscale of these magnetic fields, they will be reflected and scattered extensively in the intermediate lenghthscale, this will mean that the 
average size of the density inhomogeneities generated by these coherent structures will be 
given by this intermediate lengthscale. 

\section{Summary and Conclusions}
In summary, we have probed the coherent structures in an axisymmetric magnetic field. In 
astrophysical plasmas the magnetic field fluctuations are often highly localized. In 
such local regions, magnetic fluctuations lead to coherent structures which change the way
particles diffuse in the plasma. The SO(3) formalism has been developed previously to 
understand these coherent structure. In the SO(3) formalism, it is possible to look at the 
higher modes of fluctuations independent of each other. We have studied the second and 
fourth order fluctuations of two different distributions and found the anisotropic 
structures for both of them. One is the standard Gaussian field while the other is a 
heavy tailed distribution. For the heavy tailed distribution, we have chosen the multivariate Cauchy 
distribution. We find that the scattering of particles will be different for the two distributions with significantly different lengthscales involved. While the Gaussian 
distribution gives us only two major lengthscales corresponding to the diffusion of a 
particle, the Cauchy distribution gives us three major lengthscales. This means that 
particle scattering and diffusion during turbulence will be different in a heavy tailed 
magnetic field. While for a Gaussian field, there is little chance of the charged particles 
being trapped or accelerated to very high energies, a magnetic field having a Cauchy distribution 
will lead to accelerated particles and charged particle trapping. We feel that detailed simulations would yield further differences between these two distributions.    

In this paper we have not considered the different cases of weak and strong turbulence. There are studies 
that discuss particle acceleration in strong turbulence in the presence of coherent structures \cite{lemoine}. They have shown that the acceleration depends on the gradient of the velocity flow of the magnetic field lines. As we see in our study, the acceleration will also depend on the distribution of the magnetic field fluctuations. Therefore a deeper study into these structures and their interaction with charged 
particles will yield a better understanding of plasma astrophysics and the physics of cosmic ray transport.

\begin{center}
 Acknowledgments
 \end{center} 
 
The authors acknowledge infrastructure support from the DST-SERB/ANRF Power Grant no.SPG/2021/002228.
SS acknowledges discussions with Dilip Kumar.


\begin{thebibliography}{100}

\bibitem{grasso} D. Grasso, H. R. Rubinstein, Phys. Rept. 348, 163–266 (2001).
\bibitem{subramanian} K. Subramanian, Rep. Prog. Phys. 79, 076901 (2016).
\bibitem{arad1}I. Arad, L. Biferale, I. Mazzitelli and I. Procaccia, Phys. Rev. Lett. 82, 5040, (1999).
\bibitem{rubinstein} R. Rubinstein, S. Kurien, and C. Cambon, Journal of turbulence, 16, 11, 1058,  (2015).
\bibitem{biferale} L. Biferale, D. Lohse, I. M. Mazzitelli and F. Toschi, J. Fluid Mech. vol.452, pp. 39–59 (2002).
\bibitem{staicu}A. Staicu, B. Vorselaars, and W. van de Water, Physical Review E 68, 046303 (2003).
\bibitem{arad}I. Arad, V. S. L’vov, and I. Procaccia, Physical Review E, 59
Pg 6753, (1999).
\bibitem{lemoine} M. Lemoine, Phys. Rev. D 104, 063020, (2021).
\bibitem{davis} Z. Davis, L. Comisso, C. Haggerty, and J. Nättilä, ApJ 999 261 (2026)
\bibitem{vlahos}Loukas Vlahos, arXiv:2604.17119v1
\bibitem{vinogradov} A. Vinogradov, O. Alexandrova, P. Démoulin, A. Artemyev, M. Maksimovic, et al. The Astrophysical Journal, 971, 88, (2024).
\bibitem{tsagas}Christos G Tsagas and Roy Maartens,  Class. Quantum Grav. 17 2215 (2000). 
\bibitem{kurkiS}H. Kurki-Suonio, M.B. Aufderheide, F. Graziani, G.J. Mathews, B. Banerjee, S.M. Chitre, D.N. Schramm, Physics Letters B, 289, Issues 3–4, Pages 211-216, (1992).
\bibitem{sanyal}S. Sanyal,  Phys. Rev. D 67, 074009 (2003).
\bibitem{das}P. K. Das, S. Sau, A. Saha, and S. Sanyal, Eur. Phys. J. C 81, 816 (2021).
\bibitem{cambon} C. Cambon and L. Jacquin, J. Fluid. Mech. 202, 295-317 (1989).
\bibitem{tautz} R. C. Tautz and S. Shalchi,  
Physics of Plasmas, Volume 17, Issue 12, article id. 122313, (2010).
\bibitem{bifarale}L. Biferale and I. Procaccia, Physics Reports 414, 43 – 164, (2005).
\bibitem{carretti}E. Carretti, et. al.  Monthly Notices of the Royal Astronomical Society, Volume 512, Issue 1,  Pages 945–959, (2022).
\bibitem{sumandey} S. Dey and G. Sigl, Astroparticle Physics 173,103153, (2025).
\bibitem{braithewaite}J. Braithewaite, Mon. Not. R. Astron. Soc. 397, 763–774 (2009).

\bibitem{zadorozhna} L.V. Zadorozhna, B.I. Hnatyk, Y.A. Sitenko,  Ukr. J. Phys.  Vol. 58, No. 4, 398, 2013.
\bibitem{lubke} J. Lübke, et al. Journal of Plasma Physics, 91(5), p. E147.(2025).
\bibitem{shalchi} A. Shalchi and A. Dosch, Phys. Rev. D 79, 083001, (2009). 
\bibitem{shalchiArendt}A. Shalchi and V. Arendt, Brazilian Journal of Physics, 54:126 (2024).
\bibitem{Ntourmousi}E. Ntormousi, L. Vlahos, A. Konstantinou and H. Isliker, Astronomy \& Astrophysics, 691, A149, (2024).
\bibitem{krtous}P. Krtous, Physical Review D {\bf 74} 065006 (2006).
\bibitem{sovan}S. Sau and S. Sanyal, Eur. Phys. J. C 80  2, 152, (2020).
\bibitem{bisht}D. Bisht, D. Kumar, S. Nayak and S. Sanyal,  International Journal of Modern Physics D  35:03 (2026).

\bibitem{abhijit} A. Ray and S. Sanyal, Physics Letters B 726  83–87, (2013)
\bibitem{papatha} V. Papathanasiou, Statistics \& Probability Letters,
Volume 7, Issue 1, Pages 29-33, (1988). 
\bibitem{kheruntsyan} K. V. Kheruntsyan et. al. Phys. Rev. Lett. 108, 260401 (2012).



\end{thebibliography}
\end{document}